# 3D Human Mesh Construction Leveraging Wi-Fi

Yichao Wang*
Florida State University
Tallahassee, FL, USA
ywang@cs.fsu.edu

Yili Ren*
Florida State University
Tallahassee, FL, USA
ren@cs.fsu.edu

Yingying Chen
Rutgers University
New Brunswick, NJ, USA
yingche@scarletmail.rutgers.edu

Jie Yang
Florida State University
Tallahassee, FL, USA
jie.yang@cs.fsu.edu

## ABSTRACT

In this paper, we present, Wi-Mesh, a WiFi vision-based 3D human mesh construction system. Our system leverages the advances of WiFi to visualize the shape and deformations of the human body for 3D mesh construction. In particular, it leverages multiple transmitting and receiving antennas on WiFi devices to estimate two-dimensional angle of arrival (2D AoA) of the WiFi signal reflections to enable WiFi devices to "see" the physical environment as we humans do. It then extracts only the images of the human body from the physical environment, and leverages deep learning models to digitize the extracted human body into 3D mesh representation. Experimental evaluation under various indoor environments shows that Wi-Mesh achieves an average vertices location error of 2.81cm and joint position error of 2.4cm, which is comparable to the systems that utilize specialized and dedicated hardware. The proposed system has the advantage of re-using the WiFi devices that already exist in the environment for potential mass adoption. It can also work in non-line of sight (NLoS), poor lighting conditions, and baggy clothes, where the camera-based systems do not work well.

## CCS CONCEPTS

• **Human-centered computing** → **Ubiquitous and mobile computing systems and tools**.

## KEYWORDS

WiFi Sensing, 3D Human Mesh, Channel State Information (CSI), Deep Learning

**ACM Reference Format:**

---

*Both authors contributed equally to this research.

## 1 INTRODUCTION

Recent years have witnessed tremendous progress in 3D mesh construction for human bodies and in adapting 3D human mesh in various emerging applications. Indeed, 3D human mesh parameterizes the 3D surface of the human body, which represents how individuals vary in height, weight, somatotype, body proportions, and how the 3D surface deforms with articulation. It describes the fine-grained 3D human body shape as well as the human poses and activities. 3D human mesh construction thus has been increasingly involved in the applications such as VR/AR content creation [27], virtual try-on [6], and exercise monitoring [35]. It is also a fundamental building block for various downstream tasks, such as animation [5], clothed human reconstruction [11], and rendering [45].

Traditional approaches for 3D human mesh construction primarily rely on computer vision technique that requires the installation of optical cameras in the environment, or wearable technique that requires dedicated sensors worn by human subjects. These approaches, however, require either significant infrastructure installation or diligent usage of wearable devices [17]. In addition, the computer vision-based systems cannot work well in NLoS or poor lighting conditions [19]. They also incur large errors when subjects wear baggy clothes [16]. Recently, Radio Frequency (RF) sensing offers an appealing alternative. It analyzes the reflections of RF signals off the human body for human activity sensing. As the RF signal can traverse occlusions/clothes and illuminate the human body, the RF-based approach works well under NLoS, poor lighting conditions, and baggy clothes. It also does not require a user to wear any dedicated sensors. As such, two RF-based systems have been proposed to construct 3D human mesh by using either FMCW RADAR or mmWave RADAR [56, 60]. These systems, however, rely on specialized and dedicated hardware as well as the RF signals that are specifically designed for providing accurate ranging or spatial shape of objects. These systems thus are less attractive to consumer-oriented use or mass adaptation due to their high cost.

In this paper, we ask whether it's possible to re-use commodity WiFi, originally designed for communication, to construct 3D human mesh for potential mass adaptation in smart environments. Earlier work has shown that the commodity WiFi is able to classify a set of pre-defined human body activities [50] and detect subtle movements, such as vital signs [24]. Recently, systems like WiPose [15] and Winect [32] are proposed to track more detailed 3D human poses for either pre-defined or free-form activities. However, none of these systems are able to provide fine-grained 3D human mesh that consists of thousands of vertices, which is several orders of magnitude larger than the number of body joints defined in 3D poses or activities. In addition, prior systems mainly feed the amplitude/phase or the Doppler frequency shifts of the WiFi signals into deep learning models for activity or pose tracking. They would require more quality training data and deeper neural networks to ensure better system robustness, since their input signal metrics are sensitive to several factors, such as the propagation environments, WiFi devices, and various people and activities. This limitation renders them less practical across different environments and for unseen people.

In our work, we demonstrate that the commodity WiFi can be leveraged to construct 3D human mesh, which has not been possible before. In particular, we propose a WiFi vision-based approach for 3D human mesh construction across different environments and for unseen people. We leverage the advances in WiFi technology to help WiFi devices "see" and visualize the human body as we humans do. Our system, Wi-Mesh, helps WiFi devices see a person by leveraging the fairly large number of antennas on the next-generation WiFi devices. Indeed, the new generation of WiFi 6 or 7 supports up to 8 or 16 antennas [8], respectively. These spatially distributed antennas at the WiFi receiver can be used to disentangle the signal reflections from the 3D surface of the human body, enabling the WiFi device to visualize the body shape and deformations. To realize such an approach, we propose to estimate the 2D AoA of the WiFi signal reflections off the human body. In particular, we propose to estimate incident angles of the signal received at the WiFi receiver in the azimuth-elevation plane, where azimuth is an angular measurement of the horizon direction for the received signal, and elevation is the angular measurement of the same signal in the vertical direction. Given the intensity and the corresponding signal reflected from each direction, we could derive a visualization or a 2D AoA image of the human body together with the static objects in the environment, similar to a gray-scale image captured by a camera. In addition to diversities of receiving antennas and subcarriers used in SpotFi [20], we further leverage the spatial diversity of transmitting antennas and the time diversity of WiFi packets to improve the spatial resolution of 2D AoA estimation. Unlike SpotFi only estimates two-dimensional information (i.e., azimuth and time of flight), in our work, the combination of all the diversities and joint estimation of four-dimensional information (i.e., azimuth, elevation, time of flight, and angle of departure) of WiFi signals significantly improve the resolution [55] and thus offer a better illustration of the shape and deformation of the human body for 3D human mesh construction.

The estimated 2D AoA image, however, contains information on both the human body and the surrounding static objects, such as walls and furniture in the same environment. We propose to eliminate reflections off static objects and focus on only the human body, making the 3D human mesh construction independent of the environment. As the reflections from the static objects don't change over time, we propose to extract the human body by subtracting the static components in consecutive 2D AoA estimation frames to eliminate the static reflections. Furthermore, we address the issue of human body specularity, in which WiFi signals may be reflected towards or away from the receiver depending on the surface orientation of different body parts. As a result, a single 2D AoA image can only capture a subset of the human body, but missing the parts that deflect the WiFi signals away from the receiver. To address this issue, we incorporate multiple 2D AoA estimations over time to capture a complete picture of the human body. Specifically, we design our deep learning model to focus on multiple consecutive 2D AoA estimations that involve the signal reflected from different parts of the human body.

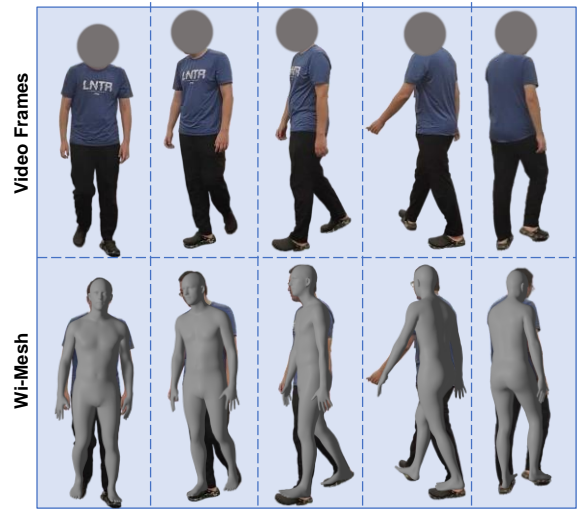

**Figure 1: 3D human mesh estimation using WiFi signals.**

For 3D human mesh construction, we leverage the most popular Skinned Multi-Person Linear (SMPL) model [25], which factors the 3D surface of the human body into thousands of vertices to represent the shape and pose deformations. To fit the SMPL model, we design deep learning models to extract both the spatial body shape and temporal body deformations from 2D AoA images. In particular, our designed deep learning models include the convolutional neural network (CNN), Gated Recurrent Unit (GRU), and self-attention mechanism. The CNN is utilized to extract the static spatial information of the human body (e.g., spatial body shape), whereas the GRU is used to analyze dynamic deformations of the human body under various poses. And the self-attention mechanism is leveraged to dynamically learn the contributions of each frame and highlight the important frames in the final representation. Finally, both the extracted spatial body shape and temporal body deformations are fitted into the SMPL model to obtain the 3D mesh representation of the human body.

We evaluate the Wi-Mesh system with twenty people in various indoor environments including the home, classroom, and laboratory. We perform 3D mesh construction across different environments, as well as for unseen people or the same people with different activities. We also compare our system to the prior WiFi-based approaches that leverage raw CSI measurements or Doppler frequency shift of WiFi signals. Figure 1 shows one example of our constructed 3D human meshes when the subject is walking and turning around. The first row shows the human body captured by the video camera, whereas the second row illustrates how the constructed 3D human meshes match the ground truth. As we can see that the generated dynamic 3D human meshes align very well with the subject, demonstrating the effectiveness of our system. The main contributions of our work are summarized as follows.

- We propose a WiFi vision-based approach for 3D human mesh construction. As the WiFi signal can traverse occlusions and clothes, our system works well under NLoS, poor lighting conditions, and baggy clothes. It also does not require a user to wear any dedicated sensors.
- We estimate the 2D AoA of the signal reflections off the human body to visualize the body shape and deformations. We then digitize the extracted human body images into 3D SMPL model by leveraging the deep learning techniques.
- Experimental results show that our Wi-Mesh system is highly accurate and robust across different environments and for unseen people, which is comparable to prior specialized and dedicated hardware-based systems. Our system has the advantage of reusing WiFi devices for potential mass adaptation.

## 2 RELATED WORK

We divide the techniques for 3D human mesh construction into three categories: computer vision-based, RF-based, and wearable sensor-based approaches.

## 2.1 Computer Vision-based Approach

With the development of deep learning algorithms and annotated datasets, 3D human mesh construction is a popular topic in the computer vision community. There exist systems that utilize only a single image to construct 3D human mesh. For example, some researchers [2, 37] made pioneer work to reconstruct 3D human mesh by leveraging silhouette and joint information in an image by optimizing a statistical body model. More recently, Kanazawa et al. [16] presented an adversarial learning framework to recover 3D human mesh directly from image pixels. Whereas Bogo et al. [7] proposed a fully automatic method to fit the 2D joints in the image to the SMPL model [25]. There are also approaches that utilize video to recover 3D human mesh. In the early stage, Hogg et al. [12] mapped a simple 3D human mesh to a walking person based on video frames. Some later systems [34] constructed the 3D human mesh by leveraging multi-view videos. For example, Tung et al. [46] proposed a method to predict SMPL parameters by using optical flow, silhouettes, and joints from video frames. However, computer vision-based approaches have fundamental limitations. They cannot work well in NLoS or poor lighting conditions, and incur large errors when subjects wear baggy clothes [16, 19].

## 2.2 RF-based Approach

Recently, many RF sensing systems have been explored to predict 3D human mesh [38]. For example, RF-Avatar presented by Zhao et al. [60] is a sensing system that utilizes FMCW RADAR [1] to infer 3D human mesh based on an adversarial training network. Xue et al. proposed mmMesh [56], which constructed the dynamic 3D human mesh by using point clouds data directly exported from the mmWave RADAR. However, these systems all require dedicated and specialized hardware, and thus are not scalable for a large number of users due to their high cost. Moreover, Huang *et al.* [13] did primipary work to image a simple object using specialized WiFi devices. Particularly, they require the use of the customized device of USRP. And their image resolution is too low to visualize the 3D surface of the human body as no spatial or frequency diversity was exploited. On the other hand, there exists work re-using commodity WiFi devices that already exist in the environment to build various sensing applications, such as large-scale human activities recognition [41, 50, 51, 54, 58], small-scale human motion detection [39, 40], vital sign monitoring [23, 24], indoor localization [57, 61], person identification [47], and object sensing [31, 42]. Recent highly related work explored commodity WiFi to estimate 2D or 3D human pose [15, 32, 33, 48]. For example, Wang et al. [48] estimate 2D human pose by using deep learning as a purely data-driven black-box solver. Moreover, WiPose [15] is proposed to track 3D pose of a set of pre-defined activities, whereas Winect [32] is developed to estimate 3D pose for free-form activities. These systems, however, only track the locations of a few body joints (less than 20), which are too cores-grained to generate the 3D human mesh that consists of thousands of vertices.

## 2.3 Wearable Sensor-based Approach

Wearable sensors have been widely using for human sensing [49, 52, 53]. Specifically, wearable sensor-based 3D human mesh construction systems require the user to wear dedicated sensors. For instance, Tautges et al. [43] reconstruct the human pose by utilizing the calibrated readings of four 3D accelerators attached to the arms and legs. Some other 3D human mesh estimation systems [14, 26] reconstruct full-body motions with the help of inertial measurement units (IMUs) sensors attached to different positions of the body. Additionally, Kaufmann et al. [17] fitted the measurements of electromagnetic sensors attached to the human body to generate the 3D human mesh. However, wearable sensor-based systems are cumbersome and incur non-negligible costs [17].

## 3 PRELIMINARY

### 3.1 WiFi Sensing

The applications of WiFi are rapidly evolving, and WiFi can do more than just communicate. Indeed, researchers have developed techniques to make sense of WiFi signals, particularly in sensing human activity. As WiFi signals travel through the air, the human body and static objects will reflect, diffract, or absorb the signal energy, which results in reflections, diffraction, and scattering, creating multipath propagation. The received signals thus carry information about human activity and the surrounding environment. Earlier work primarily focuses on utilizing Received Signal Strength (RSS), which is a single quantity per packet that represents the Signal to Interference & Noise Ratio (SINR) over the entire channel bandwidth, for device-based localization [22] or device-free intrusion detection [59]. However, RSS is a coarse-grained metric. Only limited information can be inferred regarding human activity based on RSS.

Recently, researchers shift the focus to CSI provided by IEEE 802.11 WiFi devices for human sensing. CSI contains amplitude and phase measurements for each orthogonal frequency-division multiplexing (OFDM) subcarrier. For example, it includes amplitude and phase for each of the 56 OFDM subcarriers on a standard 20MHz channel. Due to frequency diversity, different subcarriers experience different multipath fading. While such effects are often averaged out for RSS measurement, the amplitude and phase of individual subcarrier are more likely to change when small movements have altered the multipath environment. The CSI exported from commodity WiFi devices thus provides fine-grained information to characterize the human activity and the multipath environment. Researchers have devoted significant efforts to build WiFi sensing applications leveraging CSI, such as large-scale activity recognition [29, 50, 54], small-scale motion detection [3, 24, 39], and object sensing [31, 42].

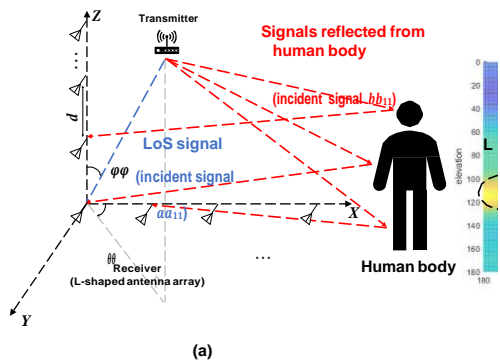

Figure 2: WiFi vision based on

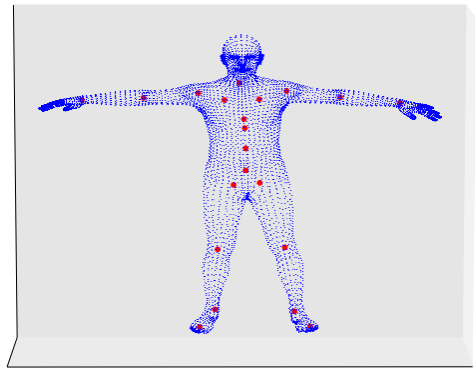

Figure 3: Vertices and joints of SMPL model.

Existing work in WiFi sensing mainly uses a black-box approach by directly inputting the CSI or the Doppler frequency shifts into deep learning models for activity recognition and tracking [15, 48, 50]. It is based on the assumption that similar activities will reflect, diffract, or absorb the signals similarly, resulting in similar signal change patterns. However, the CSI or its Doppler frequency shifts are more or less sensitive to different multipath environments, heterogeneous WiFi devices, and various people and activities. These systems normally work well for pre-defined activities in pre-trained multipath environments, but require more training data and deeper neural networks to have better system robustness. This largely limits their applicability across different environments and for a diverse set of people and activities. In addition, CSI or the Doppler frequency shifts cannot directly provide any spatial information about the physical environments (e.g., the shape of the human body or static objects) as that of the FMCW RADAR or mmWave RADAR. Existing WiFi sensing systems thus cannot be directly used to construct fine-grained 3D human mesh, which requires the 3D spatial information of the human body and poses.

### 3.2 2D AoA-based WiFi Vision

In our work, we propose a concept of 2D AoA-based WiFi vision, which leverages the advances in WiFi technology to help WiFi devices "see" and visualize the physical environment as we humans do. We estimate the 2D AoA of the WiFi signal reflections to enable WiFi devices to visualize the shape and deformations of the human body for 3D mesh construction. In particular, the new generation of WiFi 6 devices can support up to 8 antennas, whereas it is increased to 16 antennas for WiFi 7 [8]. These spatially distributed antennas can be used to separate the signal reflections from different directions/locations, providing spatial information about the physical environment. Thus, the 2D AoA of the WiFi signal reflections provides spatial information of the objects that reflect the WiFi signals and can be used to visualize the shape and the poses of the human body, similar to a gray-scale image captured by a camera.

We illustrate our idea with Figure 2, where an L-shaped antenna array (i.e., $N$ antennas) is used to receive the WiFi signal transmitted from the transmitter. We note that the L-shaped antenna array is among the best to estimate 2D AoA [44]. In Figure 2, the L-shaped antenna array is aligned with the X-Z axis. As WiFi signals travel through space, they will be reflected from the human body as well as from other static objects (e.g., walls and furniture) in the environment. We can calculate the 2D AoA of the signal reflections based on the phase shift of the received signals at multiple antennas. We denote the phase shift of the received signal on the $n^{th}$ antenna as:

$$\Phi_n(\varphi, \theta) = e^{-j2\pi[X\sin(\varphi)\cos(\theta)+Z\cos(\varphi)]/c}, \quad (1)$$

where $\varphi$ is the elevation angle, $\theta$ is the azimuth angle, and $X$ and $Z$ are coordination of the $n^{th}$ antenna corresponding to the reference antenna at the origin of coordinates, $d$ is the distance between two adjacent antennas, and $c$ is the speed of light. Next, we can obtain the phase shifts across the $N$ antennas in the antenna array as follow:

$$\mathbf{a}(\varphi, \theta) = [\Phi_1(\varphi, \theta), \Phi_2(\varphi, \theta), ..., \Phi_N(\varphi, \theta)]^T, \quad (2)$$

where **a** is called the steering vector. With the steering vector, the 2D AoA of the signal reflections can be derived by using the MUSIC algorithm [36]. And we refer to the estimated 2D AoA spectrum as a 2D AoA image in our work.

Figure 2(b) shows an example of the estimated 2D AoA image. The X-axis shows the azimuth direction, whereas the Y-axis represents the elevation direction. The color dots in the figure illustrate the intensity of the signal reflected from corresponding azimuth-elevation directions. From Figure 2(b), we can observe the direction of the LoS (i.e., with the strongest intensity) and the spatial information of the human body that reflects WiFi signals. This serves as the basis of our proposed WiFi vision-based 3D human mesh construction. However, since $N$ antennas can only resolve up to $N-1$ distinct signal reflections. With 8 to 16 antennas on next-generation WiFi devices, the corresponding 2D AoA spatial resolution is still insufficient to describe the surface of the human body (e.g., head, torso, and limbs). Thus, we need to further improve the spatial resolution in order to construct the 3D human mesh.

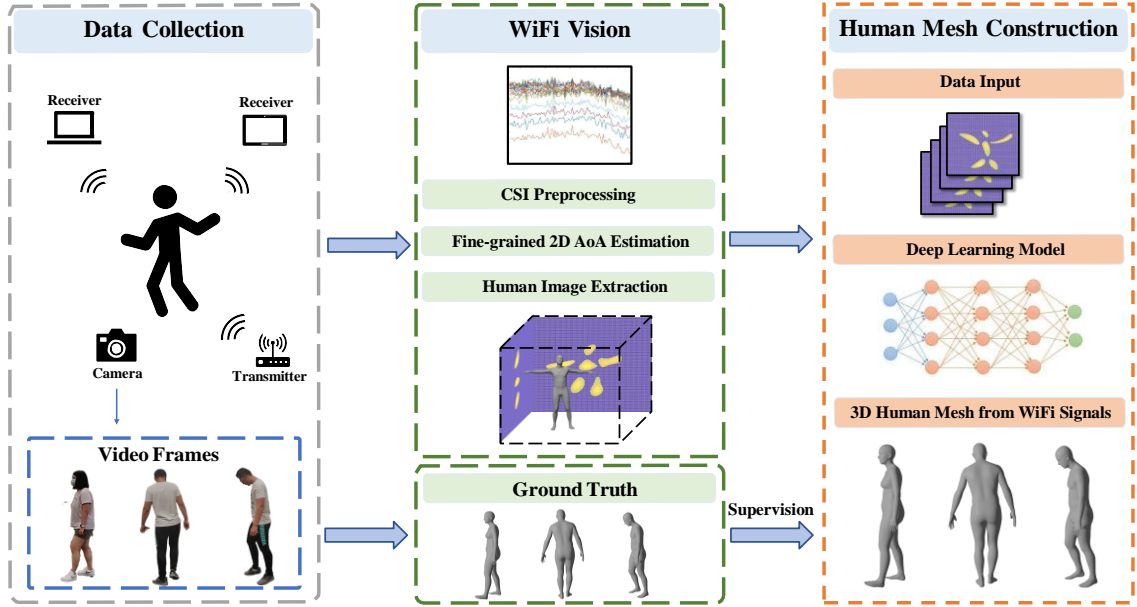

**Figure 4: Wi-Mesh system overview.**

## 3.3 SMPL Model

In this work, we adopt the SMPL model [25] to represent the 3D human mesh. SMPL is a parametric statistical 3D human body model to encode the 3D mesh into two low-dimensional parameters, which are shape and pose parameters. The shape parameter $\beta$ represents the top $m$ coefficients of a Principal Component Analysis (PCA) shape space, which controls the statistical shape features of the subject and is independent of various poses. The pose parameter $\gamma$ is used to control changes in the joints of the human body, which is represented by one global 3D rotation vector and relative 3D rotation of 23 joints in axis-angle representation. SMPL model provides an efficient mapping method $M(\beta, \gamma)$ that digitizes the shape and pose parameters into 6890 vertices that fully characterize the 3D surface and poses of the human body.

Figure 3 illustrates one example of 3D point cloud of the SMPL model, which includes 6890 vertices and 23 joints. In particular, the big red dots are these 23 joints, which are used to represent the motions or poses of the human body. The thousands of small blue dots are the surface vertices, which describe the 3D surface of the human body, and can be used to represent the height, weight, somatotype, and body proportions of the human body. Compared with 3D human pose [15, 32] that only consists of a few (e.g., less than 15) body joints, a 3D human mesh contains thousands of vertices, which are more complicated and require spatial information of the human body. Therefore, the prior WiFi-based activity and 3D pose tracking systems cannot be directly leveraged to construct accurate 3D human mesh.

## 4 SYSTEM DESIGN

### 4.1 System Overview

The key idea of our work is to leverage the advances of WiFi and 2D AoA estimation of the signal reflections to visualize the shape and deformations of the human body for 3D mesh construction. As illustrated in Figure 4, the system takes as input time-series CSI measurements at multiple antennas of two WiFi receivers when a person is moving around in the environment. The system can reuse existing WiFi devices and take advantage of CSI measurements from existing traffic or the system-generated probing packets for measurement purposes. The WiFi signals reflected from different parts of the human body and surrounding objects will arrive at the receiver in various directions (i.e., azimuth and elevation angles).

The CSI measurements then go through preprocessing to remove the random phase offsets. Then, our system estimates the 2D AoA of the signals reflected from the human body and static objects. In contrast to existing work, SpotFi [20], which only uses spatial diversity of receiving antennas and frequency diversity of subcarriers, we further leverage spatial diversity of transmitting antennas and time diversity of the WiFi packets to jointly estimate four-dimensional information instead of two-dimensional information in SpotFi and thus greatly improve the resolution of the 2D AoA estimation. Next, our system conducts human image extraction to filter out the signals reflected by the static objects in the environment (e.g., walls and furniture) and only focus on only the human subject. As each 2D AoA image can only capture a subset of the human body due to human body specularity, we further combine multiple 2D AoA images to have a full picture of the human body.

Next, we design a deep learning model to extract both spatial information and temporal deformation of the human body from the 2D AoA images to construct the 3D human mesh. The deep learning model has three components: CNN, GRU, and the self-attention module. Among them, CNN is used to parse the static spatial information of the whole human body. The GRU is utilized to extract the dynamic deformation of the body in the temporal dimension. And the self-attention mechanism is used to adaptively learn the contributions of features and highlight the important ones in the final representation. Finally, the extracted spatial body shape and temporal body deformations are fitted into the SMPL model to obtain the 3D mesh representation.

Our system could benefit from the prevalence of WiFi signals and re-use the WiFi devices that already exist in the environment for potential mass adoption. It thus presents tremendous cost savings when compared with dedicated and specialized hardware-based systems. In addition, as the WiFi signals can traverse occlusions/clothes and can illuminate the human body, our system can work under NLoS, poor lighting conditions, or baggy clothes, where the camera-based systems do not work well.

### 4.2 CSI Preprocessing

Due to the hardware imperfection of the commodity WiFi device, the CSI measurements suffer random phase offsets incurred by sampling time offset (STO) and packet detection delay (PDD) across packets. As the AoA information will be derived from the phase shifts at multiple antennas of the receiver, we first need to perform CSI preprocessing to remove the CSI random phase offsets. Specifically, we adopt a linear fit method proposed in [20] to sanitize the random phase shift. The optimal linear fit method is described as follows:

$$\sigma = argmin \sum_{k,v=1}^{K,V} \sum_{y=1}^{Y} (\Upsilon(k,v,y) + 2\pi f_\delta(y-1)\alpha + \beta)^2, \quad (3)$$

where $f_\delta$ is the frequency difference of the adjacent OFDM subcarriers, $\Upsilon(k,v,y)$ represents the unwrapped phase of the CSI at the $k^{th}$ subcarrier of a packet, which is transmitted from the $v^{th}$ transmitting antenna and received at the $y^{th}$ receiving antenna, $\alpha$ is the common slope of the received phase responses for all antennas, and $\beta$ is the offset. The $\sigma$ includes the time delay of each WiFi packet. Finally, we can get the calibrated CSI phase $\bar{\Upsilon}(k,v,y)$ by removing the time delay as $\bar{\Upsilon}(k,v,y) = \Upsilon(k,v,y) - 2\pi f_\delta(y-1)\sigma$.

### 4.3 Fine-grained 2D AoA Estimation

Besides the spatial diversity of receiving antennas and the frequency diversity of OFDM subcarriers utilized in the previous work (i.e., SpotFi [20]), our work further leverages the spatial diversity of transmitting antennas and time diversity of WiFi packets to achieve the fine-grained 2D AoA estimation. Specifically, SpotFi utilizes 3 receiving antennas and 30 subcarriers, which result in 90 sensing elements in total. In our work, we utilize 3 transmitting antennas, 9 receiving antennas, and 30 subcarriers, which result in 810 sensing elements on each receiver. In addition, we leverage the spatial diversity of receivers by placing two receivers at different locations which can capture the environment and human body from different angles. Thus, the number of sensing elements in our system is an order of magnitude larger than that of SpotFi. Also, the time domain WiFi packets can provide time diversity which helps to optimize the 2D AoA estimation.

Moreover, the previous work, SpotFi, applies the 2D MUSIC algorithm to estimate only two-dimensional information including the azimuth and the time of flight (ToF). In our work, we further extend the MUSIC algorithm to the 4D MUSIC algorithm which can estimate four-dimensional information including azimuth, elevation, ToF, and angle of departure (AoD). Thus, the resolution can be significantly improved by jointly exploiting information from four signal dimensions [55].

In particular, the 2D AoA estimation is to leverage the phase shifts of the received signals at spatially separated antennas of the L-shaped antenna array of the WiFi receiver. First, the spatially separated antennas at the receiver will observe phase shifts for the signal reflected from each direction. Moreover, the spatially separated transmitting antennas on the transmitter will also introduce the phase shifts. In addition, different OFMD subcarriers will cause phase shifts for the same reflection as well due to frequency differences. Multiple consecutive packets can also provide time diversity. All of these diversities can be incorporated to improve the resolution of the 2D AoA estimation.

First, we incorporate the phase shifts introduced by the spatial diversity of transmitting antennas. We describe the phase shift $\Psi(\omega)$ across transmitting antennas as follows:

$$\Psi(\omega) = e^{-j2\pi f d \sin(\omega)/c}, \quad (4)$$

where $\omega$ is AoD of the signal, $d$ is the distance between two adjacent transmitting antennas, and $f$ is the frequency of the signal.

Next, we integrate the phase shifts associated with the frequency diversity of the OFDM subcarriers. For evenly distributed subcarriers, the phase shift across two adjacent subcarriers can be described as follow:

$$\Omega(\tau_n) = e^{-j2\pi f_\delta \tau_n/c}, \quad (5)$$

where $\tau_n$ is the ToF of the $n^{th}$ propagation path.

Then, we leverage CSI measurements of the WiFi signals across all subcarriers of all the receiving and transmitting antenna pairs to build a sensing elements array on each receiver. Compared with Equation 2 in Section 3.2, the new steering vector can be denoted as:

$$\bar{a}(\theta, \varphi, \tau) = [1, ..., \Omega_\tau^{M-1}, \Phi_{(\theta,\varphi)}, ..., \Omega_\tau^{M-1}\Phi_{(\theta,\varphi)}, ..., \Phi_{(\theta,\varphi)}^{N-1},$$
$$..., \Omega_\tau^{M-1}\Phi_{(\theta,\varphi)}^{N-1}]^T, \quad (6)$$

$$\mathbf{a}(\theta, \varphi, \omega, \tau) = [\bar{a}(\theta, \varphi, \tau), \Psi_\omega \bar{a}(\theta, \varphi, \tau), ..., \Psi_\omega^{K-1}\bar{a}(\theta, \varphi, \tau)]^T, \quad (7)$$

where $\mathbf{a}(\theta, \varphi, \omega, \tau)$ is the steering vector formulated by phase difference across each of the sensing elements, $\Omega_\tau$, $\Psi_\omega$ and $\Phi_{(\theta,\varphi)}$ are the abbreviations of $\Omega(\tau)$, $\Psi(\omega)$ and $\Phi(\theta, \varphi)$.

After incorporating the spatial and frequency diversities to construct the sensing elements, we next combine multiple consecutive WiFi packets (i.e., time diversity) to further improve the quality of 2D AoA estimation. On the contrary, SpotFi merely utilizes one WiFi packet to estimate the AoA. The intuition is that the estimation variance for the covariance matrix in the MUSIC algorithm will decrease as more WiFi packets are used, resulting in sharper peaks in the AoA spectrum. This improves the quality of the 2D AoA image as it is easier to differentiate different signal reflections from different body parts or subjects. In our work, we combine multiple WiFi packets to generate thirty 2D AoA spectrums per second to synchronize with the sampling rate of the vision-based ground truth.

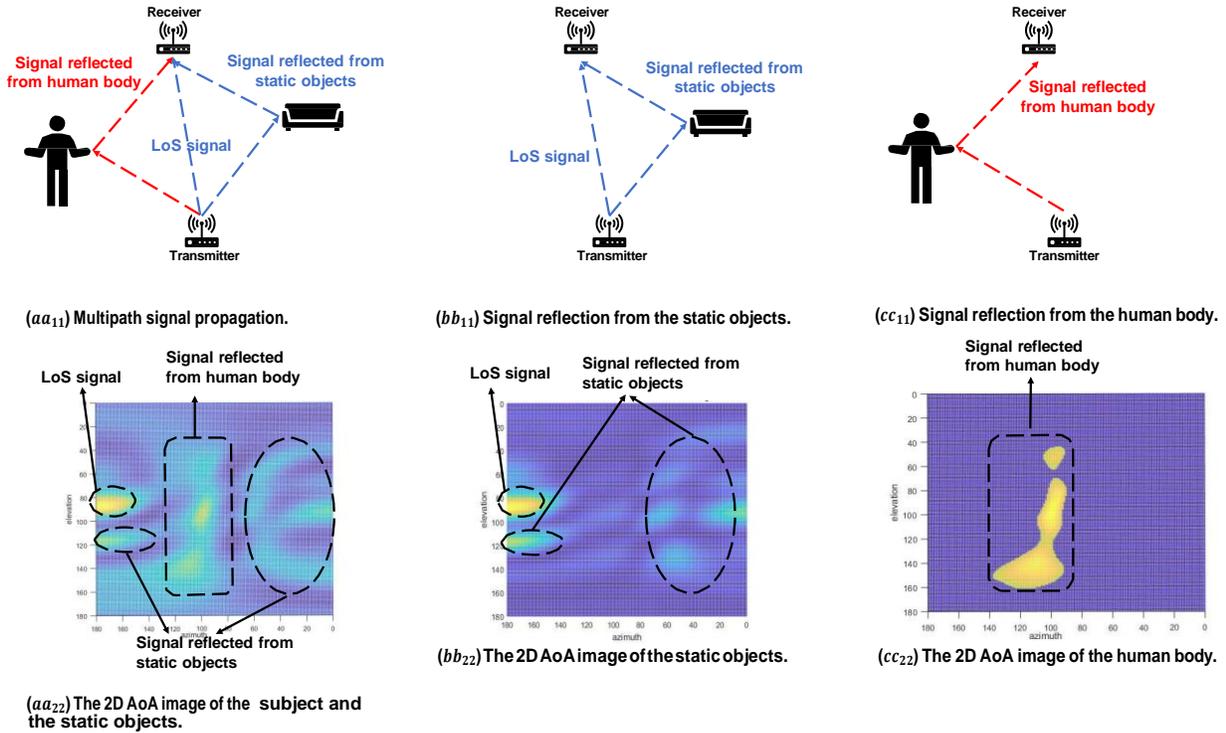

Figure 5: Illustration of human motion extraction.

Unlike SpotFi that only utilizes 2D MUSIC to estimate azimuth and ToF, we extend it into a 4D MUSIC algorithm which can achieve joint estimation for parameters of azimuth, elevation, AoD, and ToF as increasing the dimension improves the resolution significantly [55]. The estimation is presented as follows by maximizing the spatial spectrum function:

$$P(\theta, \varphi, \omega, \tau) = \frac{1}{A^H(\theta, \varphi, \omega, \tau) E_N E_N^H A(\theta, \varphi, \omega, \tau)}. \quad (8)$$

Finally, we accumulate the 2D AoA values in the ToF and AoD dimensions to generate fine-grained 2D AoA spectrums, which contain the spatial information of both the human body and the surrounding static objects.

### 4.4 Human Image Extraction

As the estimated 2D AoA contains the information about both the human body and the surrounding static objects (i.e., LoS propagation from the transmitter, walls, and furniture), we need to extract only the human body for 3D mesh construction by removing the reflections off static objects and LoS propagation. Figure 5 shows the LoS and multipath propagation when the subject is performing activities. We can observe LoS signals, signal reflections from static objects, and signals bounced off the human body. The corresponding 2D AoA estimation is shown in Figure 5, from which we can differentiate various signal reflections or LoS signals based on their incident angles. Moreover, we can observe that compared to the LoS signals, the signals reflected from the human body are relatively weaker. Eliminating the LoS signals and the signal reflections from static objects could make our system independent of different multipath environments or heterogeneous WiFi devices.

Since the reflections from the static objects remain the same over time, we can extract the human body and eliminate the static reflections by subtracting the static components in consecutive 2D AoA estimation frames. For example, we show the signal reflections of the static objects in Figure 5, and the corresponding 2D AoA estimation in Figure 5. As these objects are static, their 2D AoA estimation does not change over time. We refer to this as the static component. We could thus estimate such a static component in consecutive 2D AoA images and subtract it to eliminate the reflections from the static objects.

After eliminating the signal reflections from static objects, we obtain the reflections mainly from the human body. This is because some human body reflections will bounce off the static objects again and then reach the receiver. We call these bounce-off signals indirect reflections. In contrast to the signal directly reflected from the human body, the indirect reflections are very weak as they propagate longer paths and experience more attenuation. Although they cannot be fully removed, we can mitigate the impact of the indirect reflections by filtering out weak reflections with an adaptive threshold. Figure 5 shows an environment with human body reflection only, whereas Figure 5 illustrates the 2D AoA of the human body after we extract human body reflections and perform indirect reflection mitigation. From Figure 5, we can observe a rough shape of a human body.

As we all know that the human body reflects WiFi signals. But the human body is considered specular with respect to WiFi signals because the wavelength of the WiFi signal is much larger than the granularity of the surface of the human body. As a result, when the human body is probed by WiFi signals, some parts of the human body may reflect the signals directly to the receiver, whereas other parts may scatter the signals away from the receiver. Therefore, a single frame of 2D AoA image may only capture a subset of the human body, but misses other parts that disperse the signals away. As shown in Figure 6, the first frame only captures the head and legs of the human body, while the second frame only has information about the middle part of the torso and arms. To address this issue, we can assemble a sequence of 2D AoA spectrums to recover all parts of the human body. The last frame in Figure 6 is an intuitive example showing the full picture of the human body after combining the previous two frames. In our system, we leverage the deep learning model to learn such aggregation of multiple frames to handle the problem of the specularity of the human body.

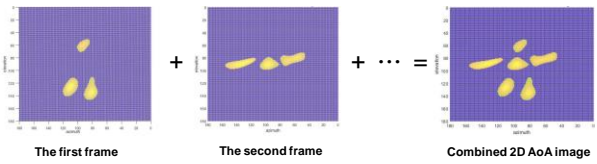

**Figure 6: Combining multiple frames**

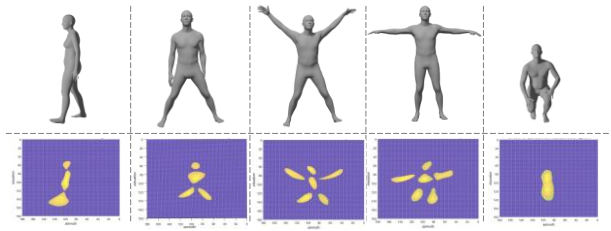

**Figure 7: Some example 2D AoA images related to different poses and shapes of the human body.**

Figure 7 shows some examples of the 2D AoA images extracted for the human body. The first row shows the ground truth of the human body and the poses, whereas the second row shows the extracted human body from the 2D AoA spectrums. We can observe that the extracted human body images match the ground truth. For example, we can see the silhouettes of the human body and differentiate different poses. Since one 2D AoA image can only describe the 2D spatial information, we will need at least 2D AoA images at two different angles to recover the 3D human mesh. We thus leverage two receivers at two different viewpoints in our system to construct the 3D mesh of the human body, as shown in Figure 8.

## 4.5 Deep Learning Model

*4.5.1 Tensor Input.* We transfer the 2D AoA images to the 4D tensors with the size of $15 \times 2 \times 180 \times 180$ as the input of our deep learning model. Specifically, the first dimension is the number of frames, the second dimension is the number of receivers, the third and fourth dimensions are the ranges of the elevation and azimuth angles of the 2D AoA image. As we combine 2D AoA images of multiple frames and receivers, the 4D tensor possesses enough spatial and dynamic information to enable the deep learning network to estimate the 3D human mesh.

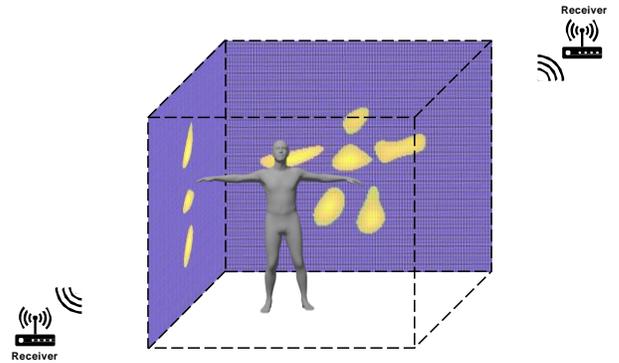

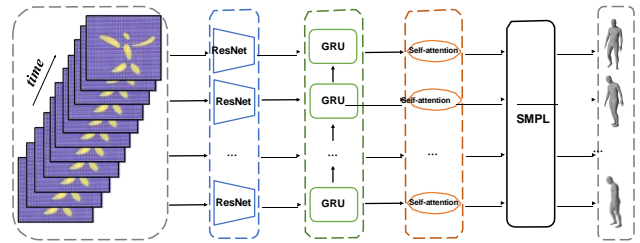

**Figure 9: Deep learning model.**

*4.5.2 Model Structure.* Our deep learning model is shown in Figure 9. The first stage of our model is a CNN-based feature extractor which can extract spatial features of each frame. Specifically, we use CNN to learn the human silhouettes and human shapes in 2D AoA images. After the CNN layer, a max-pooling operation is used to remove redundant information and focus on the most relevant features. Then, a high-level vector of human body representation can be derived.

After the feature extraction, we need to find an appropriate way to tackle the problem about specularity of the human body as aforementioned, and also relate the representation vectors of each frame. When the subject is performing dynamic motions, there exist high temporal dependencies between the consecutive frames since human motion changes are continuous and the next motion is closely related to the previous motion. And the past information can help fix and constrain the current motion state. Therefore, we feed a sequence of the above-extracted representation vectors to a multi-layer GRU to model the temporal coherence of the sequence and infer the missing parts based on previous frames. The recurrent networks update their hidden states as they process data in sequence. Then, we concatenate the hidden states corresponding to each frame and output a sequence elements. After that, a dropout layer is encoded to prevent the model fall into the local minimum. However, each previous frame has varying degrees of influence on the current result over time toward dynamic motions. Thus, we implement the self-attention mechanism to dynamically learn the contributions of each frame in the sequence elements, assign corresponding weights to the hidden state at each frame for highlighting the contribution of the important ones, and generate the final representation. After getting the output from the self-attention layer, we further map the result vector into the shape and pose representation of the SMPL model by utilizing a multi-layer fully connected neural network. The length of the result vector is 82, including 72 pose vectors and 10 shape vectors. Then, we input the shape and pose vectors to the gender-neutral SMPL model [25] that consists of 6890 vertices to represent the 3D human mesh.

*4.5.3 Model Loss.* We encode the SMPL model in our system to represent the 3D human mesh, which utilizes shape and pose parameters as input and outputs a 3D human mesh consisting of thousands of vertices. Thus, our problem can be considered a regression problem that regresses the shape and poses parameters of the SMPL model. In our work, the loss function has two components: the shape losses $L_\beta$ and pose losses $L_\gamma$. We use the $\ell_1$ norm to evaluate differences between the predicted param e ground truth.

We leverage the standard adversarial training schemes, and the adversarial loss term that is backpropagated to our model is:

$$L = \frac{1}{T} \sum_{t=1}^{T} \|g_t - \bar{g}_t\|_{L_1}. \quad (9)$$

where $g_t$ and $\bar{g}_t$ are the predicted parameters and the corresponding ground truth, and $T$ is the number of frames in the input sequence.

The overall loss of our model is a weighted sum of pose losses and shape losses, which can be written as follow:

$$L_{TOTAL} = \alpha_J L_\gamma + \alpha_V L_\beta, \quad (10)$$

where $\alpha_J$ and $\alpha_V$ are the weights assigned to pose and shape losses, respectively.

## 5 EXPERIMENTS
### 5.1 Experimental Setup

*5.1.1 Devices.* In our experiments, we leverage the commodity WiFi devices, Dell LATITUDE laptops, as the WiFi transmitter and WiFi receivers. We deploy one WiFi transmitter and two WiFi receivers at each experimental site. Specifically, the transmitter contains three linearly-spaced antennas. And the WiFi receiver has nine antennas in L-shape. The L-shaped antenna array has two subarrays in the orthogonal direction which both consist of two Intel 5300 Network Interface Cards (NICs), as shown in Figure 10(a). We also utilize the signal splitter shown in Figure 10(b) to stitch NICs with shared antennas to simulate possible antenna configuration of the new generation WiFi devices. The cost of each Intel 5300 NIC is around ten dollars. The antennas on the receiver are equally spaced, where each antenna is half a wavelength apart (2.8 cm). The WiFi channel used is at 5.32 GHz with 40 MHz bandwidth, and the default transmitting packet rate is 1000 packets per second. We utilize Linux 802.11 CSI tools [9] to capture CSI measurements of 30 OFDM subcarriers for each packet. Also, we utilize the camera to record the ground truth of the human body and activities. In addition, we use the vision-based approach in [18] to obtain high-resolution ground truth of the pose information and utilize the VideoAvatar [4] to capture the body shape information as the ground truth. The network time protocol (NTP) is used to enable synchronization for all devices.

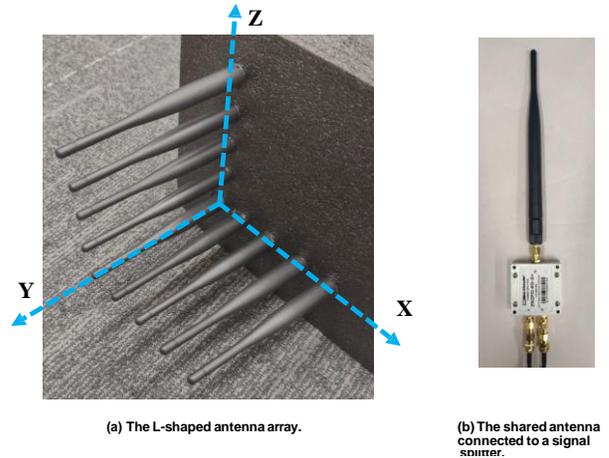

(a) The L-shaped antenna array.  (b) The shared antenna connected to a signal splitter.

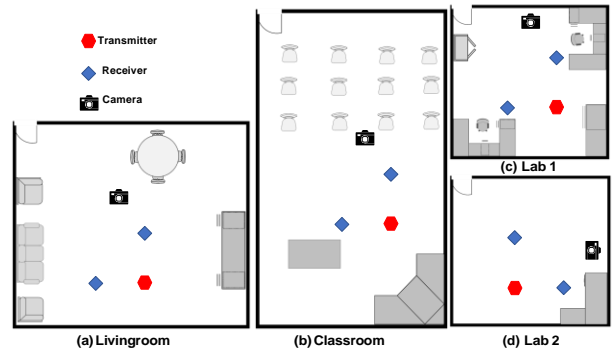

(a) Livingroom  (b) Classroom  (c) Lab 1  (d) Lab 2

**Figure 11: Layout of experiment environments.**

*5.1.2 Data Collection.* In the experiments, we recruit 20 volunteers (12 males, 8 females) of different heights and weights. Each volunteer is asked to perform everyday activities, including walking in a circle, walking in a straight line, walking with arm motions (i.e., any arm motions like lifting arms, swinging arms, waving hands, etc), lifting the legs in the place, random arm motions in the place, and rotating torso. The experiments are conducted in four different real-world environments including a classroom, two laboratories, and a living room. Figure 11 shows the detailed layout of these environments. the default distance between the transmitter and the receivers is $2m$. The size of the two laboratories is similar (i.e., at around $4.5m \times 4.5m$), but with different furniture. The sizes of the classroom and living room are $8.5m \times 5.5m$ and $6m \times 6m$, respectively. In total, we collect around sixty million WiFi CSI packets to train and test our system for 3D human mesh construction. The data collection was approved by the IRB of the authors' institution.

*5.1.3 Model Settings.* We utilize the ResNet-18 framework [10] as our feature extractor in our deep learning model, where $\epsilon_i \in R^{2048}$ is the max pooled features of the last layer. We then implement 2 layers of GRU as the residual model. The number of the hidden state is set as 2048. The dropout rate of the dropout layer is 0.5. For the self-attention module, we utilize two fully-connected layers of size 2048 and $\tanh(\cdot)$ activation.

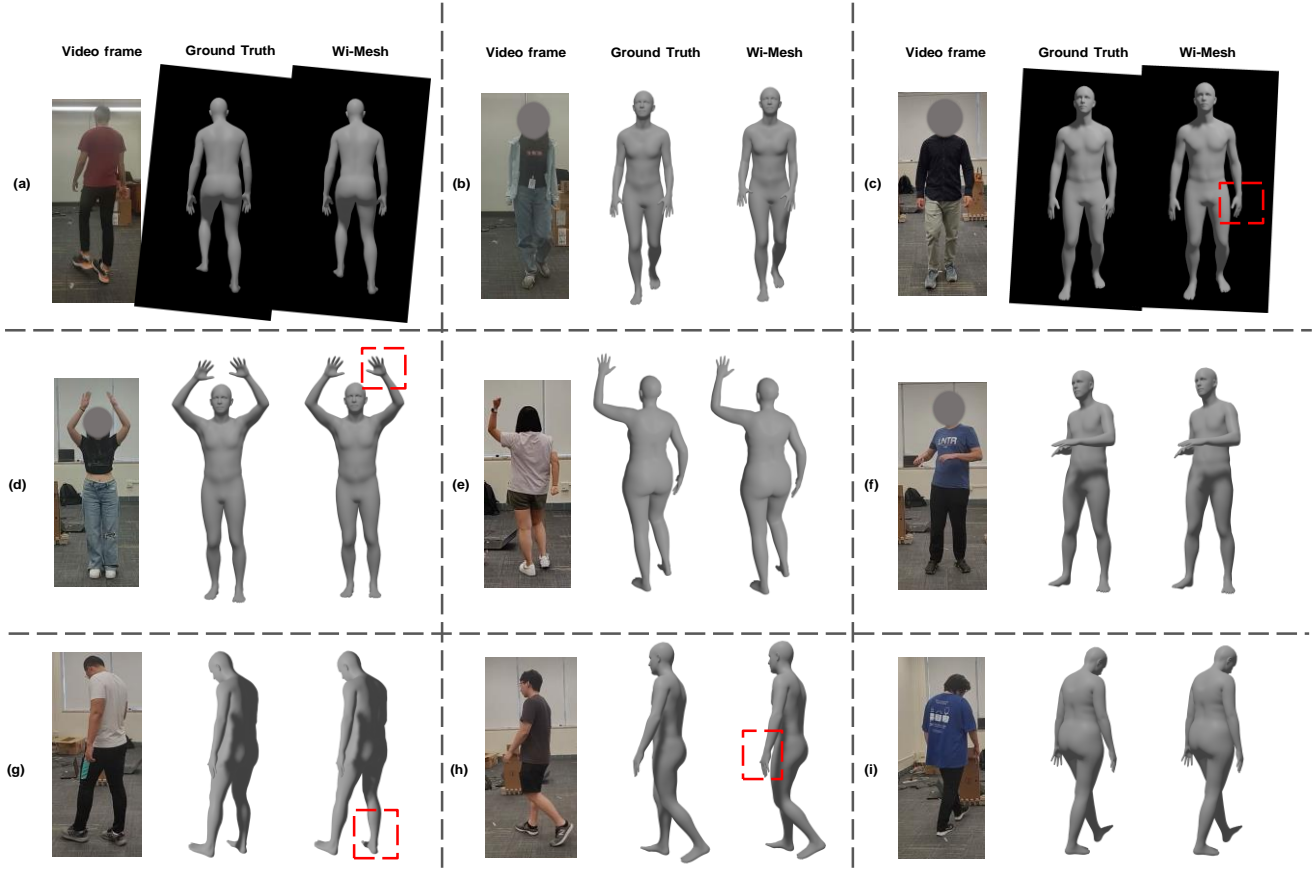

**Figure 12: The examples of the constructed 3D human mesh.**

During the training period, we split the data with 20 subjects into two non-overlapping datasets, which consist of a training set with 80% of the data, and a testing set with the rest of 20% data, making sure that the testing data is not seen by the model while training. Moreover, we conduct 5-fold cross-validation to test the robustness of our model. We set the initial learning rate to 0.0001 with periodical decay, and the batch size is set to 16. The weights assigned to the loss function are set to 1 and 0.05 respectively. We implement our deep learning network in PyTorch. And we leverage Adam optimizer to train our model on NVIDIA RTX 3090 GPU.

*5.1.4 Baselines.* To study the effectiveness of our proposed WiFi vision-based approach, we compare the results of our system with the results of using raw CSI measurements (i.e., CSI-based), Doppler frequency shift of WiFi signals (Doppler-based), and 1D AoA (1D AoA-based) as input data based on the same deep learning model of Wi-Mesh. In particular, the CSI-based baseline utilizes the CSI measurements of 30 OFDM subcarriers as input, whereas the Doppler-based baseline leverages the frequency-time Doppler profiles as system input. These two traditional input metrics are widely used in prior WiFi sensing systems [15, 48, 50]. The 1D AoA-based baseline uses two 1D AoA spectrums at each receiver (i.e., elevation and azimuth are estimated separated as the 1-dimensional metric) as the system input. It is worth noting that all of these input metrics are extracted from the same subjects and data collected in Section 5.1.2 for a fair comparison.

*5.1.5 Evaluation Metrics.* We leverage the commonly used per vertex error (PVE) and mean per joint position error (MPJPE) for our evaluation. Among them, MPJPE is the average Euclidean distance between the predicted joint locations of the human mesh and the ground truth, and PVE is the Euclidean distance between the predicted human mesh vertices and the corresponding vertices on the ground truth mesh.

## 5.2 Overall Performance

We first evaluate the overall performance of our system and compare it with other baselines. We note that the training and testing data are non-overlapping and the testing data includes unseen environments and subjects, and the same subjects under different activities. The overall results are shown in Table 1. As we can see, Wi-Mesh significantly outperforms other baselines and can estimate the vertices and joint locations of the 3D human mesh accurately. Specifically, the average PVE and MPJPE of Wi-Mesh are only 2.81$cm$ and 2.4$cm$, respectively. Among the other three baselines, the CSI-based method performs the worst as the raw CSI measurements are very sensitive to different multipath environments, the configuration of the WiFi devices (e.g., power, layout, etc.), and various people and activities. The Doppler-based method is better than the CSI-based but still has very large errors. This is because the Doppler shift measures dynamic changes in the signals, which is less sensitive to the multipath environment, but still got affected by the configuration of the WiFi devices and various people and activities. The AoA-based approaches perform better than the prior two metrics as both 1 AoA and 2 AoA can capture the intrinsic spatial information of the human body, which is independent of propagation environments, heterogeneous WiFi devices, or various people and activities. Nevertheless, the proposed 2D AoA-based approach performs the best since 2D AoA jointly estimates azimuth and elevation, and thus can uniquely identify the body shape and deformations in 2D space, whereas two 1D AoAs of the body shape or deformations experience ambiguity in 2D physical space.

**Table 1: Overall system performance comparison.**

|               | PVE (cm) | MPJPE (cm) |
|---------------|----------|------------|
| CSI-based     | 12.14    | 9.75       |
| Doppler-based | 6.01     | 4.51       |
| 1D AoA-based  | 4.10     | 3.51       |
| Wi-Mesh       | 2.81     | 2.40       |

**Table 2: System performance comparison for unseen environments.**

|               | PVE (cm) | MPJPE (cm) |
|---------------|----------|------------|
| CSI-based     | 18.67    | 14.61      |
| Doppler-based | 9.94     | 7.71       |
| 1D AoA-based  | 5.56     | 4.42       |
| Wi-Mesh       | 3.02     | 2.61       |

**Table 3: System performance for different unseen environments.**

|             | PVE (cm) | MPJPE (cm) |
|-------------|----------|------------|
| Classroom   | 2.78     | 2.36       |
| Living room | 2.92     | 2.56       |
| Lab1        | 3.12     | 2.63       |
| Lab2        | 3.04     | 2.64       |

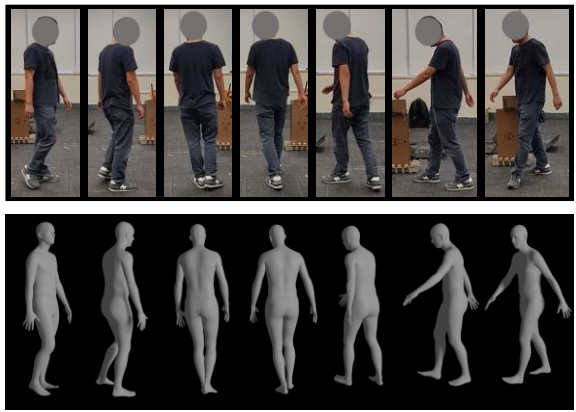

**Figure 13: The examples of dynamic 3D human meshes generated by Wi-Mesh.**

Two recent RF-based 3D human mesh estimation systems, mmMesh and RF-Avatar [56, 60], reported slightly better or similar performance when compared to Wi-Mesh. In particular, the PVE and MPJPE of mmMesh are 2.47*cm* and 2.18*cm* [56], respectively, while they are 1.65*cm* and 5.75*cm* for RF-Avatar [60], respectively. These two systems, however, utilize specialized and dedicated hardware (i.e., mmWave RADAR or FMCW RADAR) as well as the RF signals that are specifically designed for sensing. Thus, they can obtain either accurate ranging or 3D point cloud of objects for 3D mesh construction. Instead, our system leverages commodity WiFi, which is originally designed for communication. Our system thus could benefit from the WiFi devices that already exist in pervasive WiFi networks for potential mass adaptation.

To qualitatively evaluate the Wi-Mesh, we illustrate the 3D meshes generated by our system for different subjects performing various activities in Figure 12. The first column in each subfigure shows the reference video frame recorded by the camera when the subject is performing the activity. While the second column shows the ground truth of human mesh produced by the vision-based method. And the third column presents the result generated by Wi-Mesh. We use red dotted boxes to highlight the mispredicted and distorted body parts. We find that most incorrect constructions are at the ends of the arms or legs. This is because the signal reflections from these body parts are relatively weak. Nevertheless, we can easily observe that the overall constructed 3D meshes match the ground truth very well.

Figure 13 further shows a sequence of video frames that a subject is walking with arm motions. The ground truth and the meshes generated by our system are shown in the first and second columns, respectively. We can observe that our system constructs dynamic 3D human meshes over time smoothly and accurately. This is because our system could extract the temporal relationship between consecutive frames, which helps the dynamic 3D human mesh construction. The above results demonstrate that our system could construct 3D human mesh accurately under various environments, humans, and activities.

### 5.3 Impact of Unseen Environments

Adapting to different and unseen indoor environments is important for the system to be practical. We thus specifically evaluate the performance of our system under unseen environments. In particular, we first train our system in the classroom setting and then test the system in the other three unseen environments (i.e., Living room, Lab1, and Lab2). We compare the results of Wi-Mesh with the other three baselines in Table 2. We observe that the performance of the CSI-based and Doppler-based approaches decreased dramatically under unseen environments. This is because both the CSI and the Doppler shifts are sensitive to different unseen environments, such as the multipath propagations and the device configurations. For Wi-Mesh, it achieves an average vertices location error of 3.02cm and a joint position error of 2.61cm. Our system thus is robust to different unseen indoor environments. This is because our system leverages 2D AoA images of the human body to extract intrinsic features for 3D mesh construction, which is independent of different environments.

In addition, we show the performance of Wi-Mesh for each of these unseen environments in Table 3. Specifically, we collect WiFi data in three environments to train the model and then test it in the reminding unseen environment. We observe that the MPJPEs are 2.36*cm*, 2.56*cm*, 2.63*cm*, and 2.64*cm* for the classroom, living room, lab1, and lab2, respectively. The PVEs are 2.78*cm*, 2.92*cm*, 3.12*cm*, and 3.04*cm* for these four environments, respectively. We also find that the results of the lab environments are slightly worse than those in the classroom and living room. This is because the labs have much smaller space, creating more indirect reflections that could negatively affect the Signal-to-noise ratio (SNR) of the extracted 2D AoA body image. Still, our system achieves accurate 3D human mash constructions under each unseen environment, demonstrating the robustness of our system regarding environment changes or deploying in totally new environments.

Table 4: System performance comparison for unseen subjects.

|  | PVE (cm) | MPJPE (cm) |
|---|---|---|
| CSI-based | 16.12 | 13.54 |
| Doppler-based | 8.26 | 6.22 |
| 1D AoA-based | 4.92 | 3.95 |
| Wi-Mesh | 3.01 | 2.73 |

Table 5: System performance for NLoS scenario.

|  | PVE (cm) | MPJPE (cm) |
|---|---|---|
| Line-of-sight | 2.81 | 2.40 |
| Wood Panel Occluded Scenario | 3.65 | 3.08 |
| Through-wall Scenario | 4.39 | 3.97 |

### 5.4 Impact of Unseen Subjects

In practical scenes, Wi-Mesh should be able to deal with subjects that our model has never seen in the training period, which places high demands on the scalability of the system. Therefore, we also investigate the performance of our system when handling unseen subjects. Specifically, we split our dataset into two non-overlapping datasets, including 15 people in the training set and the other 5 non-overlapping people in the testing set. The results are shown in Table 4. We can see that our system can handle unseen subjects effectively with an average PVE of 3.01*cm* and MPJPE of 2.73*cm*, which dramatically outperforms the other three baselines. It is mainly because our system could leverage the 2D AoA image of the human body to directly extract features of the body shape and deformation for 3D mesh construction. The above results demonstrate the robustness of our proposed system to unseen people or activities.

### 5.5 Impact of NLoS and Baggy Clothes Scenarios

To evaluate the NLoS scenarios, we place wood panels with a thickness of 2.5*cm* between the subjects and the WiFi devices (one transmitter and two receivers). In this configuration, the WiFi signals which are emitted from the transmitter and reflected from the human body will be blocked by the wood panels. We also place the WiFi transmitter and receivers in one room and let the subjects perform activities in another adjoining room to test our system under the through-wall scenarios. The trained Wi-Mesh model under the LoS scenario is directly utilized to test the NLoS scenarios.

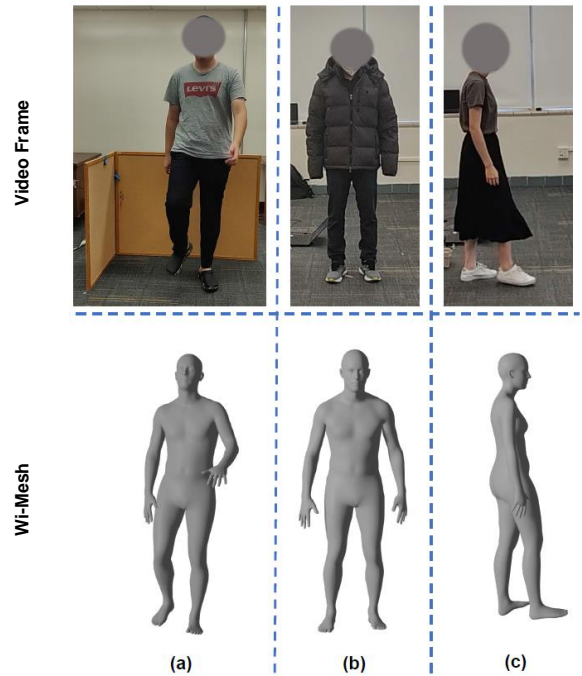

Figure 14: The examples of NLoS and baggy clothes scenarios.

Table 5 presents the results of the shape and pose estimation under both LoS and NLoS scenarios. The PVE and MPJPE are 3.65*cm* and 3.08*cm* for wood panels block scenarios, respectively. And the PVE and MPJPE are 4.39*cm* and 3.97*cm* for through-wall scenarios, respectively. In the NLoS scenarios, our system can still construct accurate 3D human mesh but performs slightly worse than that of LoS scenarios. The reason for the slight performance degradation under NLoS is that when the signal traverses the occlusions, the signal strength is attenuated, resulting in lower SNR of the WiFi signals.

Figure 14 shows some examples of the 3D mesh construction under NLoS and baggy clothes scenarios. In Figure 14(a), the WiFi devices are blocked by the wood panels placed behind the human body. Figure 14(b) and (c) show two 3D human mesh constructed when the subjects wear baggy clothes. We can observe that the constructed 3D human mesh for the blocked subject is aligned with the ground truth. In addition, our system can accurately construct the 3D human mesh when subjects wear baggy clothes. These results show that our system works well for both NLoS and baggy clothes scenarios, where the traditional computer vision-based system will not work.

### 5.6 Impact of Distance Between Transmitter and Receiver

In this subsection, we evaluate how the system performs under different distances between the transmitter and the receiver. Specifically, we evaluate different distances between the transmitter and receiver including 1.5*m*, 2.5*m*, and 3.5*m*, which can roughly produce the reflection path length corresponding to 4*m*, 6*m*, and 8*m*, respectively. In Table 6, we can observe that the PVE are 2.67*cm*, 3.15*cm*, 3.71$cm$, and MPJPE are 2.25$cm$, 2.69$cm$, 3.07$cm$ for these three reflection path lengths, respectively. It shows that the performance decreases as the reflection path length increases. This is because the longer reflection path length results in lower received signal strength, especially after the signals are reflected by the human body. Still, our system provides considerable accuracy even when the reflection path length is 8$m$, which could cover the area of a typical room. We anticipate that a smart home environment will include a variety of smart and IoT devices with WiFi interfaces. Therefore, the high density of smart and IoT devices provides us with the opportunity to achieve good system performance.

**Table 6: System performance under different distances between transmitter and receiver.**

|       | PVE (cm) | MPJPE (cm) |
|-------|----------|------------|
| 1.5$m$ | 2.67     | 2.25       |
| 2.5$m$ | 3.15     | 2.69       |
| 3.5$m$ | 3.71     | 3.07       |

**Table 7: System performance for different numbers of receivers.**

|                       | PVE (cm) | MPJPE (cm) |
|-----------------------|----------|------------|
| Two receivers system  | 2.81     | 2.40       |
| Single receiver system| 3.80     | 3.51       |

### 5.7 Impact of Number of Receivers

In this subsection, we compare our system performance utilizing two receivers and a single receiver. Specifically, we use the collected data from two receivers and one receiver to test our system separately. And the overall data process pipeline remains unchanged. The results are shown in Table 7, we can see that the PVE and MPJPE for two receivers are 2.81$cm$ and 2.40$cm$ separately. And the PVE and MPJPE are 3.80$cm$ and 3.51$cm$ for a single receiver system. The accuracy increases when two receivers are utilized. This is because two receivers at two different viewpoints provide more dimensional information than only one receiver. While our system could still have considerable performance when utilizing data from only one receiver, which demonstrates that the overall data processing method of our system is effective and efficient.

## 6 DISCUSSION

The experiment results show that Wi-Mesh achieves promising results. However, the current system still has some limitations.

**Multiple Subjects.** Our current system only supports one person in the environment. Constructing 3D human mesh for multiple persons simultaneously, especially in a crowded space (e.g., train stations and shopping malls), is still an open problem. This is also a challenging problem for camera-based systems in the computer vision community. One promising direction to address this is to first segment each person in the 2D AoA images, and then perform 3D human mesh construction one by one. Still, it is an open question of how many persons can be supported simultaneously with the proposed WiFi vision-based approach due to the limited resolution of the 2D AoA estimation.

**Sensing Range.** Although we have demonstrated our system can work at several meters range, it is still limited when compared to the communication range of the WiFi. This is because our approach relies on the signal reflections, whose signal power is several orders of magnitude weaker than that of the LoS signals. The sensing range of our approach is thus much shorter than the communication range. This limitation could be potentially addressed by leveraging the ubiquitous WiFi networks (i.e., having close WiFi devices to cover each area collaboratively), or by leveraging powerful directional antennas.

**Computation Cost.** In our work, we utilize one desktop to perform the 2D AoA estimation with one-degree angular resolution in order to obtain fine-grained 2D AoA images and achieve good results for the 3D human mesh. Thus, our system is not optimized for the computational cost or for the real-time 3D human mesh estimation. The computational cost of our 3D human mesh construction consists of two parts. The first part is the computational complexity of the 4D MUSIC algorithm and the second part is the cost of human mesh estimation using the model. We note that once the model is trained, the time to estimate the human mesh could be negligible. The majority of the computational cost is the 2D AoA estimation using the 4D MUSIC algorithm. While the eigenvalue decomposition of MUSIC has the complexity of $O(N^2J)$ (where $J$ is the number of incident wireless signals, $N$ is the number of sensors), the major computational complexity of current implementation is searching in 4D space, which is $O(M^4)$, where $M$ is the resolution at each dimension. In particular, the current resolution of azimuth and elevation in our system is 180. However, our system may reduce the computation cost from two aspects. First, we could reduce the resolution of each dimension while still maintain a good system performance. Second, we could leverage dimension reduction-based MUSIC algorithms [21] to simplify the 4D estimation problem into two separate 2D estimation problems. We could also apply low-complexity algorithms [28, 30] to infer the incident signals without searching the 4D space.

## 7 CONCLUSION

In this work, we propose a WiFi vision-based 3D human mesh construction system, Wi-Mesh, which could benefit from the prevalence of WiFi signals and reuse the WiFi devices that already exist in the environment. In particular, we exploit the advances of WiFi and 2D AoA estimation of the signal reflections to visualize the shape and deformations of the human body for 3D mesh construction. Our system leverage deep learning models to digitize the 2D AoA image of the human body into SMPL model-based 3D mesh representation. Extensive experiments demonstrate that Wi-Mesh is highly effective and robust in 3D human mesh construction across different environments and for unseen people. Our system can also work under NLoS, poor lighting conditions, or baggy clothes, where the camera-based systems do not work well.

## 8 ACKNOWLEDGMENTS

We thank the anonymous reviewers for their insightful feedback. This work was partially supported by the NSF Grants CNS1910519, CNS2131143, DGE2146354, CNS2120396, CCF1909963, and CCF2211163.